\documentclass{article} 
\usepackage{iclr2025_conference,times}
\usepackage{algpseudocode}

\usepackage{amsmath,amsfonts,bm}

\def\eqref#1{equation~\ref{#1}}

\def\1{\bm{1}}

\DeclareMathAlphabet{\mathsfit}{\encodingdefault}{\sfdefault}{m}{sl}
\SetMathAlphabet{\mathsfit}{bold}{\encodingdefault}{\sfdefault}{bx}{n}

\usepackage{float}

\usepackage{hyperref}
\usepackage{url}
\usepackage{graphicx}
\usepackage{tikz}

\usepackage[ruled,vlined]{algorithm2e}
\usepackage{amsmath}
\newcommand{\bigO}{\mathcal{O}}

\title{LIGHTHOUSE: Fast and precise distance to shoreline calculations from anywhere on earth}

\author{Patrick Beukema*, Henry Herzog, Yawen Zhang, Hunter Pitelka, Favyen Bastani \\
Allen Institute for AI (Ai2) \\
*correspondence: patrickb@allenai.org}

\iclrfinalcopy

\begin{document}
\maketitle
\begin{abstract}
We introduce a new dataset and algorithm for fast and efficient coastal distance calculations from Anywhere on Earth (AoE). Existing global coastal datasets are only available at coarse resolution (e.g. 1-4 km) which limits their utility. Publicly available satellite imagery combined with computer vision enable much higher precision. We provide a global coastline dataset at 10 meter resolution, a 100+ fold improvement in precision over existing data. To handle the computational challenge of querying at such an increased scale, we introduce a new library: Layered Iterative Geospatial Hierarchical Terrain-Oriented Unified Search Engine (Lighthouse). Lighthouse is both exceptionally fast and resource-efficient, requiring only 1 CPU and 2 GB of RAM to achieve millisecond online inference, making it well suited for real-time applications in resource-constrained environments.
\end{abstract}

\section{Introduction}
\label{sec:intro}
\begin{figure}[H]
\centering
\includegraphics[width=1.0\linewidth]{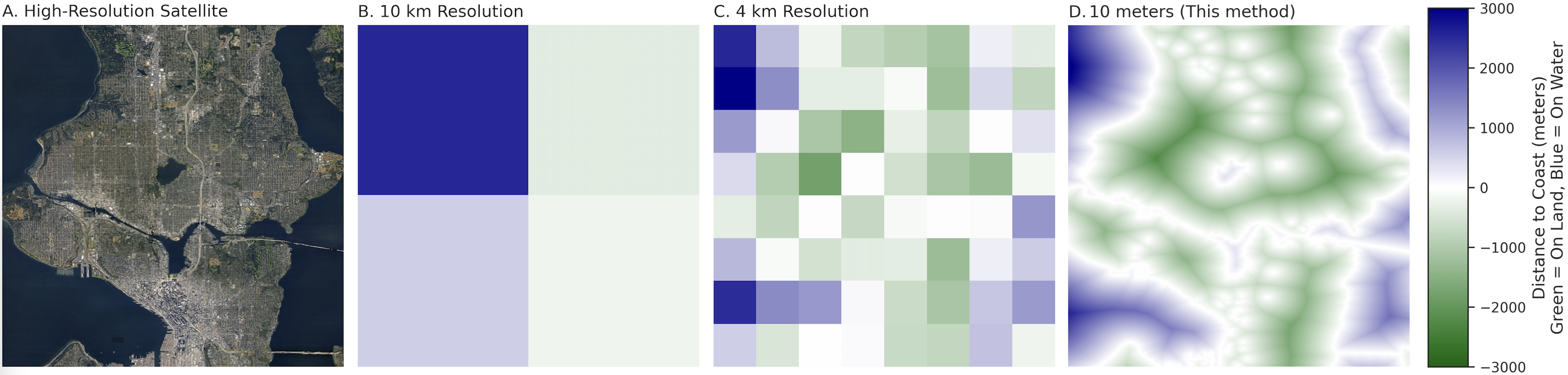}
  \caption{Distance to coast at various spatial resolutions. Location: Fremont, Seattle, Earth.}
  \label{fig:comparison}
  \vspace{-.25cm}
\end{figure}

Regularly updated and precise sea-land demarcations are crucial for tracking coastal erosion, infrastructure planning, habitat changes, and many earth-oriented machine learning models. Accurate high resolution coastal data improves both precision and recall for satellite imagery based object detection and segmentation applications and GPS modeling. For many tasks, the value of this data scales proportionally with its resolution, particularly for activities nearshore or inland (see figure: \ref{fig:comparison}). Our contributions are twofold: 
\begin{enumerate}
\item{We release a 10 meter coastal dataset including inland bodies of water.}
\item{We provide a library that efficiently generates the nearest coastal point from AoE.}
\end{enumerate}

\section{Previous work}
There are several existing publicly available global coastal distance datasets (see table \ref{tab:distance-to-coast}) spanning 2--200 km. In addition, there are a variety of global land cover maps, from which coastal points can be derived  \footnote{Besides global maps, there are a variety of higher resolution regional land cover maps. For example, NOAA has produced a 1-meter resolution land cover map covering the contiguous U.S., Hawaii, and select U.S. territories \citep{NOAA_CCAP}.}. We develop a global high resolution dataset by selectively merging World Cover ~\citep{zanaga2022esa}, a 10-meter resolution land-cover map derived from Sentinel-2 satellite images from the European Space Agency (ESA), with crowdsourced coastline annotations from OpenStreetMap. WorldCover exhibits high overall accuracy, but omitted hundreds of islands in Micronesia, Antarctica, and other areas (see supplemental fig: \ref{fig:alltiles}). OpenStreetMap exhibits lower spatial resolution globally but it covers all islands and includes recent annotations for Antarctica. Thus, merging these two datasets yields a high-quality complete global map. 

However, querying this dataset with the same methods used for 4-km resolution data is cost-prohibitive in both compute and storage. We develop a highly optimized hierarchical search algorithm (library and API) to reduce these costs. This API supports both batch queries (which benefit from caching and vectorized computations) or single instance lookups. Our approach requires only modest hardware for inference (1 CPU, 2 GB RAM) and processes distance-to-shoreline queries from AoE in milliseconds (single query: $<$ 10 ms, batch: see fig: \ref{fig:response_times} for empirical response times).

\begin{table}[H]
\caption{Existing publicly available global coastline datasets}
\label{tab:distance-to-coast}
\begin{center}
\begin{tabular}{llll}
\multicolumn{1}{c}{\textbf{Source}}  & \multicolumn{1}{c}{\textbf{Resolution (m)}} & \multicolumn{1}{c}{\textbf{Coverage}} & \multicolumn{1}{c}{\textbf{Additional Notes}} \\ \hline \\
\href{https://github.com/allenai/lighthouse}{\textbf{this method}} & $\sim 10$ & global & see section: \ref{caveats}; open source \\
\href{https://www.generic-mapping-tools.org/remote-datasets/earth-dist.html}{GSHHG} & 1855 &  high seas  & resolution: 1 arc minute  \\
\href{https://oceancolor.gsfc.nasa.gov/resources/docs/distfromcoast/}{NASA OBPG} & 4000       & high seas & also available interpolated to 1 km \\
\href{https://www.arcgis.com/apps/mapviewer/index.html?webmap=0974b50f520a43a885056e3170c85707}{ArcGIS/ESRI}  & 200000     & land only & land to high seas \\

\end{tabular}
\end{center}
\end{table}
\section{What we did}

\subsection{Dataset concatenation}
High resolution satellite imagery provides a means to generate high resolution coastlines, as the boundary between land and sea is a straightforward application of segmentation via computer vision (supervision on permanent water labels). A variety of global land cover maps have been created in the last several years including ESA's WorldCover \citep{zanaga2022esa}, Google's Dynamic World \citep{brown2022dynamic}, and ESRI, Impact Observatory, and Microsoft's LULC map \citep{karra2021global}. We selected ESA's WorldCover V2 because it has been shown to exhibit the highest accuracy for permanent water bodies \citep{xu2024comparative}. However, ESA omitted several key areas including Antarctica (unavailable at the time of their publication), along with hundreds of islands in Micronesia. We filled in the blanks with Open Street Map's crowdsourced annotations of land-sea labels \citep{osm}. We concatenated these two datasets to complete a map of the planet and resampled into 1x1 degree tiles for the purpose of parallelization and caching (see algorithm: \ref{alg:landsea}, fig. \ref{fig:alltiles}). 

\subsection{Coastal point generation}
\begin{figure}
  \centering
   \includegraphics[width=1.0\linewidth]{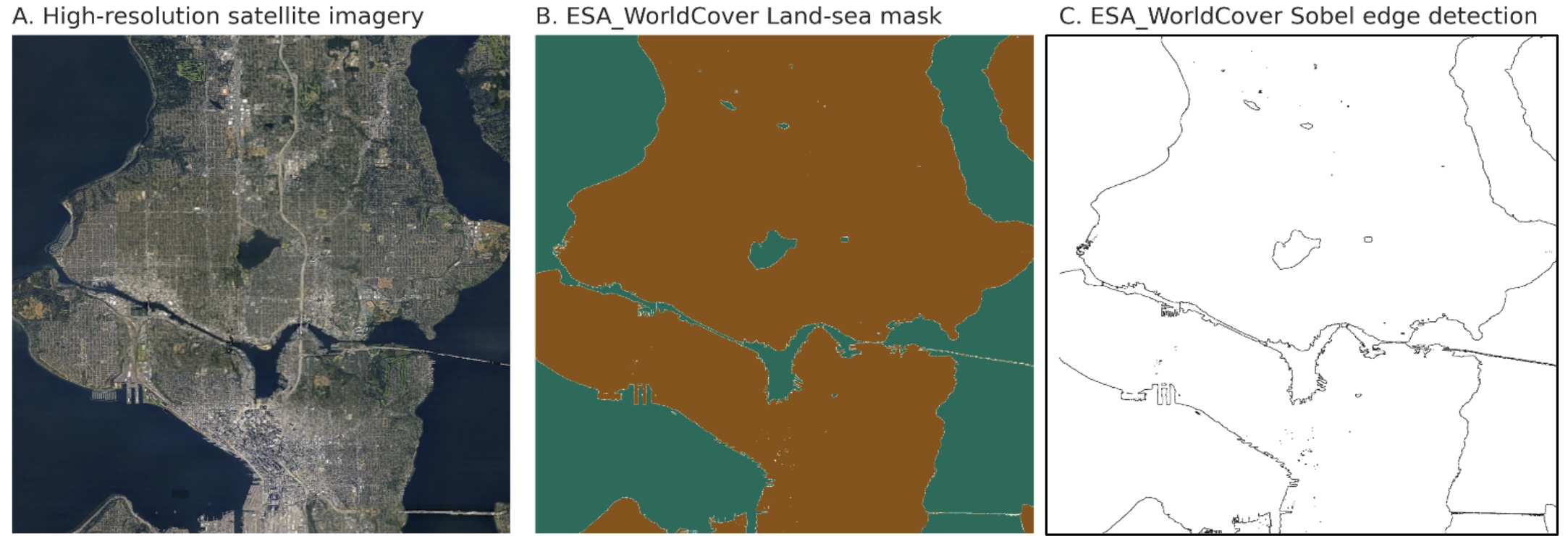}
   \caption{Visual depiction of the method. A. High resolution satellite imagery centered on Seattle. B. Binary mask constructed from sea (vs. land) labels from ESA. C. result of applying Sobel edge detection to the binary mask.}
   \label{fig:pipelines}
\end{figure}

To extract the coastal points from the joint dataset, we binarized the labels (water vs. rest), ran Sobel edge detection over the resulting mask, and then constructed balltrees for each tile using the Haversine\footnote{Note that Vincenty's formula \citep{vincenty1975direct} is more accurate than Haversine, especially over long distances, but it comes with much greater computational complexity.} metric over the coastal points (see algorithm:\ref{algorithm-edge-generate}). Throughout the codebase, we chose options that minimized latency at the expense of an increase in storage. For example, we do not compress the balltrees at all. Compression would have resulted in only a modest reduction in disk space at the expense of a significant increase in read latency. Small increases in latency can kill real-time applications and disk space is cheap in comparison to RAM and CPU. We chose h5 for the geotiffs because of the ability to query the land cover class of a single pixel without needing to load the entire file into memory. Processing was identical for both the Open Street Map tiles and ESA tiles (fig:\ref{fig:pipelines}). 

\begin{algorithm}[H]
\label{algorithm-edge-generate}
\SetAlgoLined 
\caption{Generate BallTrees of Land-Sea Edges}\label{alg:landsea}
\KwData{Open Street Map's land polygons and worldcover data}
\KwResult{BallTrees of land-sea edges for each 1x1 degree land tile, saved to disk}
Divide the world into 1x1 degree tiles that contain land\;

\For{each land tile \(T\)}{
    Generate a binary land-sea map for \(T\)\; 
    
    Identify land-sea boundaries via sobel edge detection\; 
    
    Extract edge coordinates, \(edge\_coords\)\; 
    
    Build balltree on \(edge\_coords\) using the Haversine metric\; 

    Save balltree (without compression)\;
    }
    
\end{algorithm}
\vspace{-.25cm}

\begin{figure}
\centering
\includegraphics[width=1.0\linewidth]{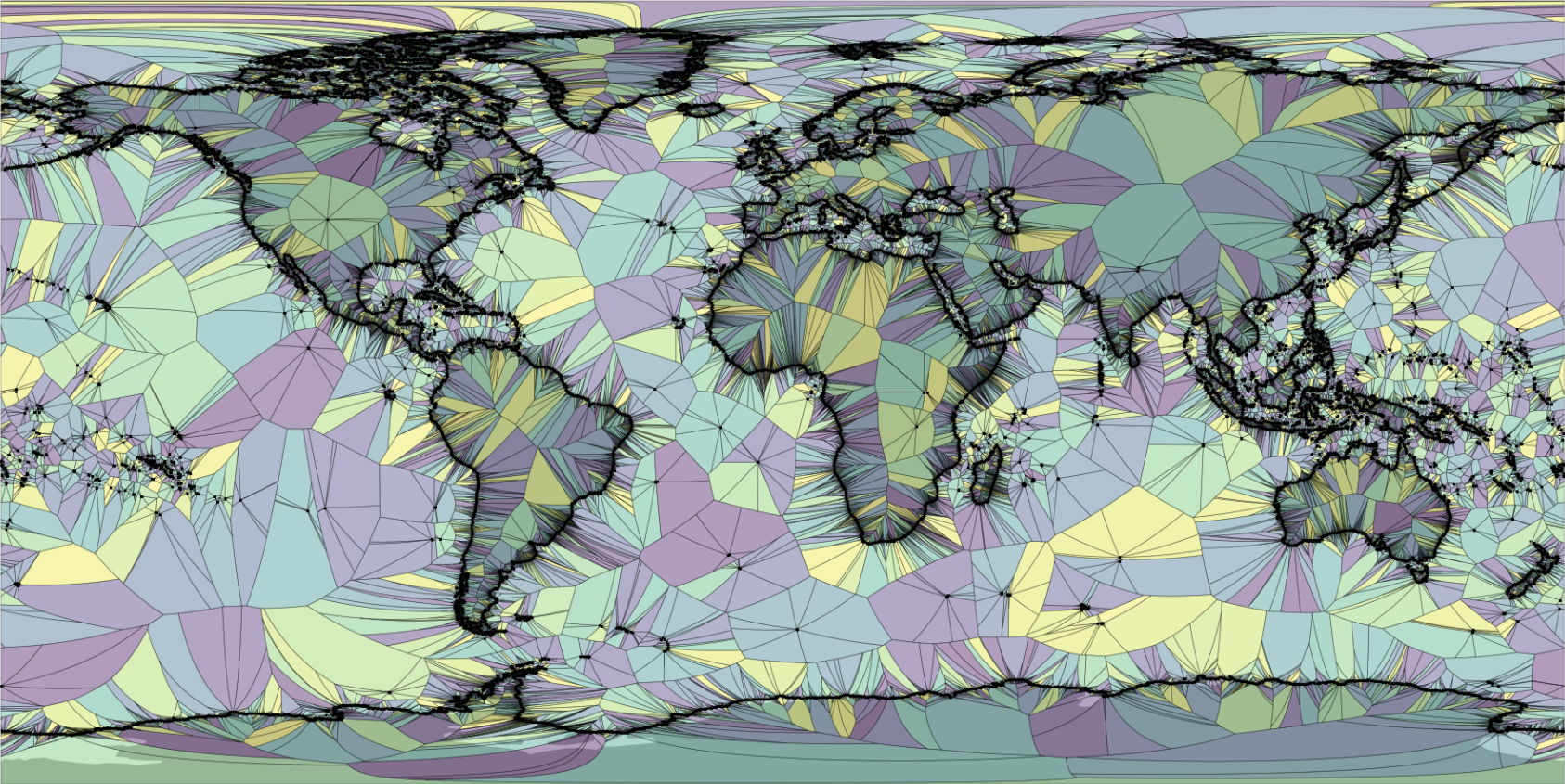}
  \caption{Spherical Voronoi tesselation of the planet based on a subset of global coastal points.}
  \label{fig:voronoi}
\vspace{-.5cm}
\end{figure}

\section{Querying the Data}
The nice thing about low resolution data is that you can precompute the distance to every point, store that data in memory, and then retrieve any point on earth in $\bigO(1)$. But that method doesn't scale to high resolution data. For example, at 10 meter resolution, one would need to store approximately 100 TB of data in RAM (float, float, int, int) which is impractical unless you have a fairly large, idling computer and do not care about money. The solution is to yield the distances at runtime, rather than caching them. To do so efficiently, for real-time applications, you need to search a large space extremely quickly.


Recall that we have high resolution ball trees for every coastal tile (fig. \ref{fig:alltiles}), and those can be queried very rapidly for the nearest coastal point. In addition to the nearest coastal point, one must also lookup the location's class label (land vs. water). Doing so rapidly requires storing the land cover maps as h5 files and retrieving just a single point's class, rather than reading the entire tile. With the ball trees and h5 files, one can immediately return the desired distance and class for any point that is contained within a tile (i.e. not on the high seas). If the point lies outside every tile, how do we determine which tile contains the nearest coastal point? The answer is via a spherical Voronoi tesselation, precomputed for the whole planet, which is also loaded in memory at runtime \citep{scipySphericalVoronoi, VanOosterom1983, caroli2010robust}. 
We generated this Voronoi tesselation (see fig. (\ref{fig:voronoi})) carefully because we could not include every point (temporal complexity scales quadratically) and if any critical point is omitted (such as an island) then the resulting tesselation and resulting distances could be incorrect. Therefore, we down-sampled the coastal points subject to the constraints that 1) every line segment in the original dataset had to be represented by at least one point in the resulting (post-resampled) dataset and 2) that the distance between connected points never exceeded a distance threshold (1km). 

\begin{algorithm}[H]
\caption{Find Nearest Land Point and Land Cover Class}
\label{alg:nearest}

Initialize an empty tile cache\;

\For{each query point $(lat, lon)$}{
    \If{$(lat, lon)$ is not in any cached tile}{
        Find and load the tile $T$ containing $(lat, lon)$ via Voronoi lookup\;
        Cache BallTree and land cover data for $T$\;
    }
    
    Use the BallTree for tile $T$ to query nearest land point and land cover class\;

    Return distance and land cover class\;
}
\end{algorithm}

\begin{figure}
  \centering
   \includegraphics[width=.75\linewidth]{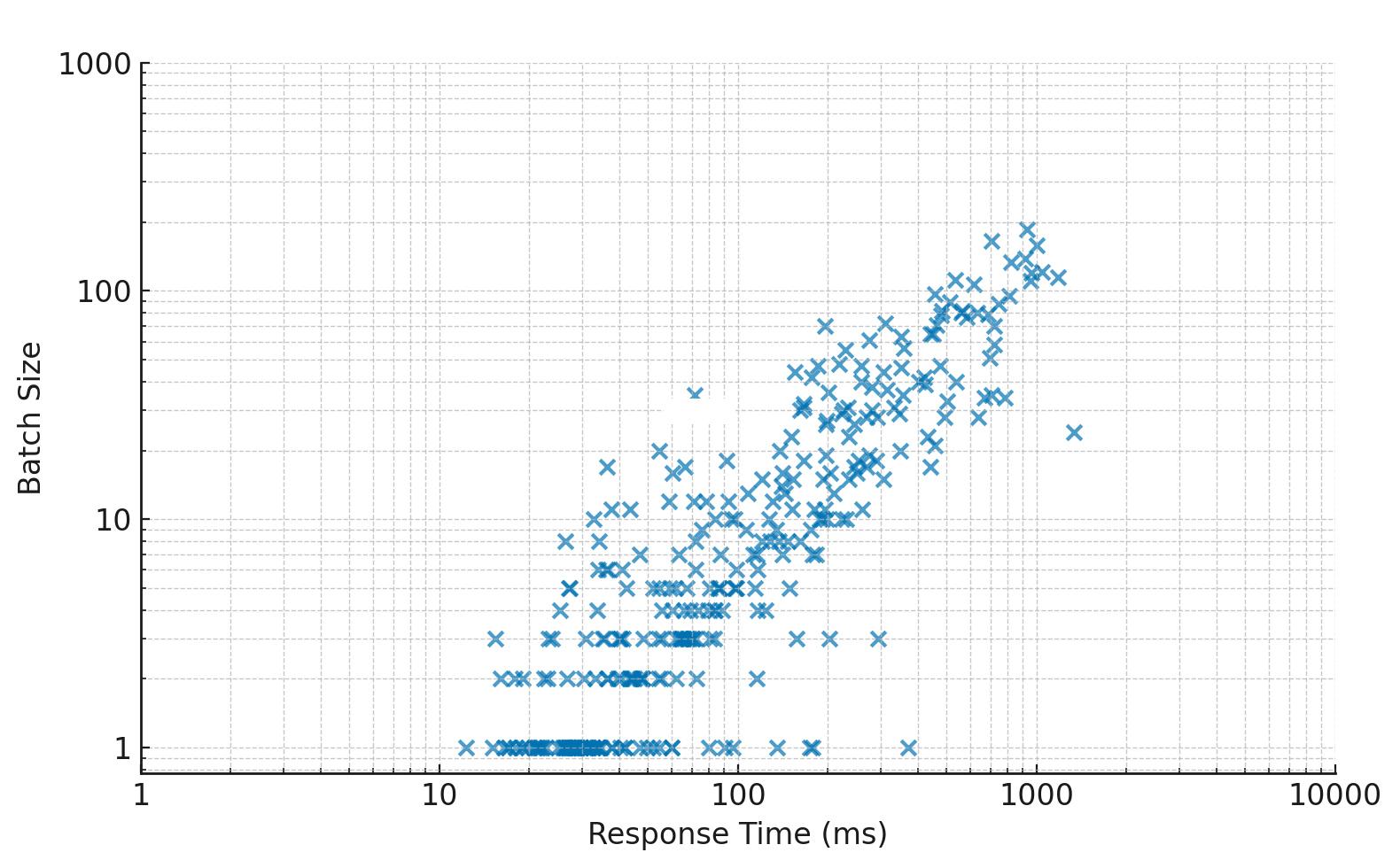}
\caption{Empirical response times from a random sample of production queries.}
   \label{fig:response_times}
\end{figure}

This method will generate distances on the fly on modest hardware at millisecond timescales. Figure \ref{fig:response_times} shows empirical latencies recorded from a random sample of batch queries from satellite imagery based vessel detections across a wide variety of requests (https://skylight.global/) in production. Response times are not quite $\bigO(1)$ like a dictionary lookup, but the resulting times are good enough for streaming real-time inference without breaking the bank.

\label{sec:caveats}
\section{Some caveats} \label{caveats}
High resolution satellite imagery -- the basis of both sets of annotations used here -- can facilitate high accuracy but it does not guarantee it. Hybridizing labels from crowd sourced maps (OSM) alongside computer vision from satellite imagery (ESA) naturally results in a mixture of errors due to human mislabeling and model misclassifications. Consider the challenges of annotating sea-land boundaries, and the complexity of cliffs, beaches, harbors, bays, wetlands, islands, etc. 

What is the definition of the coastline anyways? Nearly 50 years ago, Mandlebrot invented a new branch of mathematics, fractal geometry, in part to discuss the complexity of coastlines and the fact that their lengths are effectively infinite \citep{Mandelbrot1982}. Even today, we hotly debate the true length of a coastline. How far inland should the coastline extend? On top of the challenge of defining it, the earth is changing rapidly. Coastlines change, people build things, sea levels are rising and islands are disappearing, glaciers calve, Antarctica sea ice expands and contracts. The higher the resolution, the greater the opportunity for error.

The reported 10-meter resolution should be considered an estimate, not a definitive upper bound. The vast majority of the base annotations come from ESA’s WorldCover V2, which is derived from Sentinel-2 satellite imagery with a 10-meter pixel resolution. However, there is no single spatial resolution for the OSM labels, as they are generated from a variety of data sources using a mix of crowd-sourced and machine annotations. In some regions, the effective resolution of OSM labels may match or even exceed 10 meter resolution, while in other regions, the effective resolution may be more coarse. We analyzed OSM's effective resolution in Antarctica as a putative worst-case complexity, since Antarctica is known for coarse sampling \citep{Hormann}. We found the median inter-label segment distance between neighboring annotations to be 35 meters (see supplemental figure \ref{sup_osm_antarctica}). Note that the true resolution of OSM data globally depends on many factors including the base resolution of the satellite imagery, the human and computer vision annotation accuracy, and the complexity of the shoreline itself (which we do not know at scale). We refer the interested reader to \cite{Hormann} and \cite{Topf2013} for more detailed analyses of the quality of OSM's coastal data and its effective resolution. 

\section{Conclusion}
The coastline is complex, perhaps infinitely so \citep{Mandelbrot1982}. We introduce a means to generate and query high precision coastal data. If even higher precision is desired, this method should scale favorably up to the maximum resolution of commercial satellite imagery (10 cm as of June 2025). Not everyone needs high-resolution coastal data, but for those who do, we have open sourced both our dataset and method under permissive licenses. 
\begin{enumerate}

\item Data: gs://ai2-coastlines/v1/data (ODbL) \footnote{URL may change. Please see: https://github.com/allenai/lighthouse. }
\item Code: https://github.com/allenai/lighthouse (Apache 2.0)

\end{enumerate}

\bibliography{iclr2025_conference}

\begin{thebibliography}{13}
\providecommand{\natexlab}[1]{#1}
\providecommand{\url}[1]{\texttt{#1}}
\expandafter\ifx\csname urlstyle\endcsname\relax
  \providecommand{\doi}[1]{doi: #1}\else
  \providecommand{\doi}{doi: \begingroup \urlstyle{rm}\Url}\fi

\bibitem[Brown et~al.(2022)Brown, Brumby, Guzder-Williams, and et~al.]{brown2022dynamic}
C.F. Brown, S.P. Brumby, B.~Guzder-Williams, and et~al.
\newblock Dynamic world, near real-time global 10 m land use land cover mapping.
\newblock \emph{Scientific Data}, 9:\penalty0 251, 2022.
\newblock \doi{10.1038/s41597-022-01307-4}.
\newblock URL \url{https://doi.org/10.1038/s41597-022-01307-4}.

\bibitem[Caroli et~al.(2010)Caroli, de~Castro, Loriot, Rouiller, Teillaud, and Wormser]{caroli2010robust}
Manuel Caroli, Pedro~MM de~Castro, S{\'e}bastien Loriot, Olivier Rouiller, Monique Teillaud, and Camille Wormser.
\newblock Robust and efficient delaunay triangulations of points on or close to a sphere.
\newblock In \emph{Experimental Algorithms: 9th International Symposium, SEA 2010, Ischia Island, Naples, Italy, May 20-22, 2010. Proceedings 9}, pp.\  462--473. Springer, 2010.

\bibitem[Hormann(2013)]{Hormann}
Christoph Hormann.
\newblock Assessing the openstreetmap coastline data quality.
\newblock \url{https://www.imagico.de/map/coastline_quality_en.php}, 2013.
\newblock Accessed: 2025-02-10.

\bibitem[Karra et~al.(2021)]{karra2021global}
Kontgis Karra et~al.
\newblock Global land use/land cover with sentinel-2 and deep learning.
\newblock In \emph{IGARSS 2021-2021 IEEE International Geoscience and Remote Sensing Symposium}. IEEE, 2021.

\bibitem[Mandelbrot(1982)]{Mandelbrot1982}
Benoit~B. Mandelbrot.
\newblock \emph{The Fractal Geometry of Nature}.
\newblock W. H. Freeman, New York, 1982.

\bibitem[{NOAA, Office for Coastal Management}(2025)]{NOAA_CCAP}
{NOAA, Office for Coastal Management}.
\newblock C-cap high-resolution land cover.
\newblock \url{https://www.coast.noaa.gov/htdata/raster1/landcover/bulkdownload/hires/}, 2025.
\newblock Coastal Change Analysis Program (C-CAP) High-Resolution Land Cover. Charleston, SC: NOAA Office for Coastal Management. Accessed [Month Year].

\bibitem[{OpenStreetMap contributors}(2024)]{osm}
{OpenStreetMap contributors}.
\newblock Openstreetmap: {F}reely {E}ditable {M}ap of the {W}orld, 2024.
\newblock URL \url{https://www.openstreetmap.org}.
\newblock Accessed: 2025-01-31.

\bibitem[Topf(2013)]{Topf2013}
Jochen Topf.
\newblock State of the osm coastline.
\newblock \url{https://blog.jochentopf.com/2013-03-11-state-of-the-osm-coastline.html}, 2013.
\newblock Accessed: 2025-02-10.

\bibitem[Tyler~Reddy()]{scipySphericalVoronoi}
Edd Edmondson Nikolai Nowaczyk Joe Pitt-Francis Tyler~Reddy, Ross~Hemsley.
\newblock scipy.spatial.sphericalvoronoi.
\newblock \url{https://docs.scipy.org/doc/scipy/reference/generated/scipy.spatial.SphericalVoronoi.html}.
\newblock Accessed: 2025-02-10.

\bibitem[Van~Oosterom \& Strackee(1983)Van~Oosterom and Strackee]{VanOosterom1983}
A.~Van~Oosterom and J.~Strackee.
\newblock The solid angle of a plane triangle.
\newblock \emph{IEEE Transactions on Biomedical Engineering}, 2\penalty0 (1):\penalty0 125--126, 1983.

\bibitem[Vincenty(1975)]{vincenty1975direct}
Thaddeus Vincenty.
\newblock Direct and inverse solutions of geodesics on the ellipsoid with application of nested equations.
\newblock \emph{Survey review}, 23\penalty0 (176):\penalty0 88--93, 1975.

\bibitem[Xu et~al.(2024)Xu, Tsendbazar, Herold, de~Bruin, Koopmans, Birch, Carter, Fritz, Lesiv, Mazur, et~al.]{xu2024comparative}
Panpan Xu, Nandin-Erdene Tsendbazar, Martin Herold, Sytze de~Bruin, Myke Koopmans, Tanya Birch, Sarah Carter, Steffen Fritz, Myroslava Lesiv, Elise Mazur, et~al.
\newblock Comparative validation of recent 10 m-resolution global land cover maps.
\newblock \emph{Remote Sensing of Environment}, 311:\penalty0 114316, 2024.

\bibitem[Zanaga et~al.(2022)Zanaga, Van De~Kerchove, Daems, De~Keersmaecker, Brockmann, Kirches, Wevers, Cartus, Santoro, Fritz, Lesiv, Herold, Tsendbazar, Xu, Ramoino, and Arino]{zanaga2022esa}
D.~Zanaga, R.~Van De~Kerchove, D.~Daems, W.~De~Keersmaecker, C.~Brockmann, G.~Kirches, J.~Wevers, O.~Cartus, M.~Santoro, S.~Fritz, M.~Lesiv, M.~Herold, N.E. Tsendbazar, P.~Xu, F.~Ramoino, and O.~Arino.
\newblock {ESA WorldCover 10 m 2021 v200}, 2022.
\newblock URL \url{https://doi.org/10.5281/zenodo.7254221}.

\end{thebibliography}
\bibliographystyle{iclr2025_conference}

\newpage
\appendix
\section{Appendix}

\begin{figure}[h]

  \centering
   \includegraphics[width=1.0\linewidth]{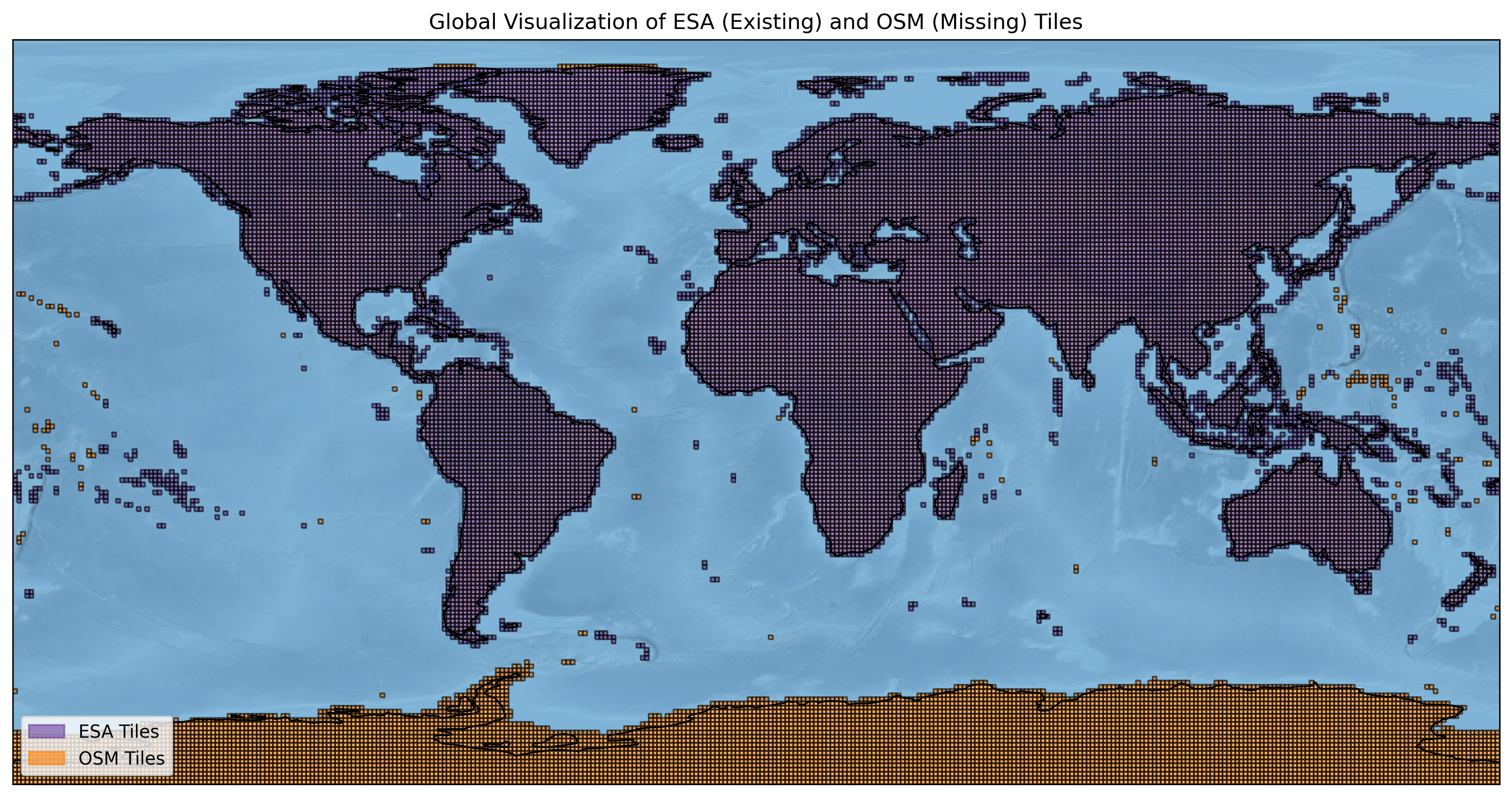}
   \caption{Distribution of 1x1 degree tiles from each data source. Note islands in Micronesia, Hawaii, South Atlantic, and Northern Greenland that were omitted in ESA's World Cover data.}
   \label{fig:alltiles}
\end{figure}

\begin{figure}[h]

  \centering
   \includegraphics[width=1.0\linewidth]{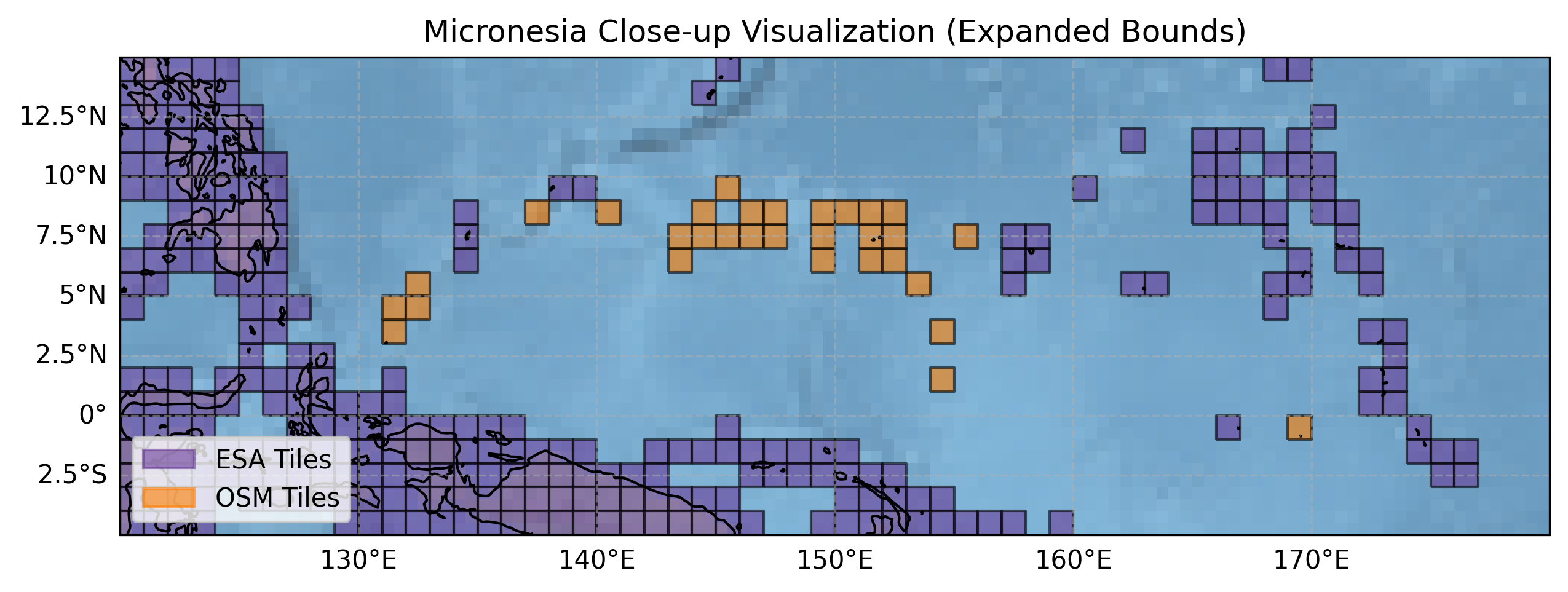}
   \caption{Close up of missing tiles from Micronesia}
   \label{fig:alltiles}
\end{figure}

\begin{figure}[h]
\label{sup_osm_antarctica}
  \centering
   \includegraphics[width=1.0\linewidth]{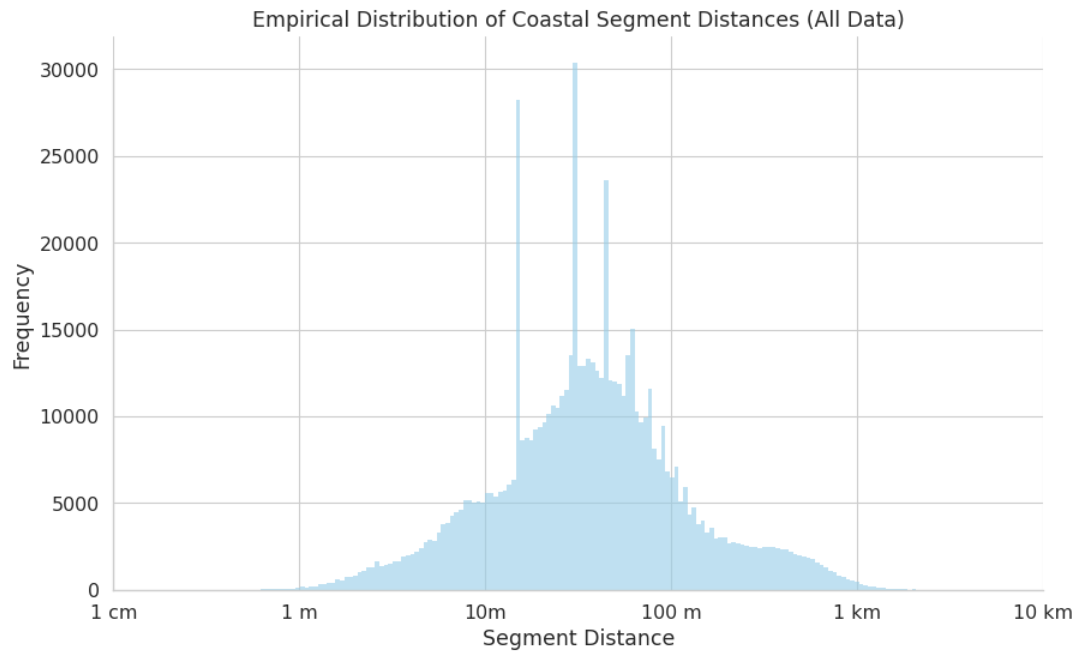}
   \caption{Empirical distribution of coastal segment distances (neighboring points) from Antarctica Open Street Map annotations}
   \label{fig:alltiles}
\end{figure}

\end{document}